\def\lte{\lower 0.5ex\hbox{${}\buildrel<\over\sim{}$}}
\def\gte{\lower 0.5ex\hbox{${}\buildrel>\over\sim{}$}}

\def\mus{\mu^{\ast}}
\def\phis{\phi^{\ast}}
\def\sg{\sigma_{\gamma\gamma}}
\def\grs{$\gamma$-rays}
\def\ee{\epsilon_1}
\def\rv{\bf r}
\def\st{\sigma_T}
\def\tg{\tau_{\gamma\gamma}}
\def\musb{\overline\mus}

\documentstyle{l-aa}
\begin{document}
\thesaurus{02.02.1 -- 02.18.5 -- 11.01.2 -- 11.14.1 -- 13.07.1}
\title {Reverberation mapping of the central regions of active galactic nuclei
using
high-energy $\gamma$-ray observations}

\author{ M. B\"ottcher \inst{1}
\and C. D. Dermer \inst{2}}

\institute{Max-Planck-Institut f\"ur Radioastronomie, Postfach 2024,
53010 Bonn, Germany
\and E. O. Hulburt Center for Space Research, Code 7653, Naval Research
Laboratory,
Washington, DC 20375-5352, USA}

\date{Recieved  ; accepted  }
\offprints{M. B\"ottcher}

\maketitle
\markboth{M. B\"ottcher, C. D. Dermer: Reverberation Mapping of AGN}{M.
B\"ottcher, C. D. Dermer: Reverberation Mapping of AGN}

\begin{abstract}

We calculate the time- and energy-dependent opacity of high-energy \grs\
attenuated by pair-production interactions with accretion-flare photons that
are scattered by gas and dust surrounding the nuclei of active  galaxies. We
show that the temporal behavior of the high-energy opacity cutoff can be
used in conjunction with the time history of the accretion flare to
determine the  location of the $\gamma$-ray emission site, as well as the
column density and scale height  of the material surrounding the central
engine. Reverberation mapping using this  technique is now possible for
nearby BL Lac objects such as Mrk 421, and will be particularly valuable
when the 10 GeV -- 100 GeV $\gamma$-ray window is opened, because here  the
attenuation of distant blazar radiation by the intergalactic infrared
background radiation is negligible.


\end{abstract}

\section{Introduction}

Markarian 421, a BL Lac object with redshift $z = 0.031$, is the first
extragalactic object detected at TeV energies with high significance (Punch
et  al. 1992). Searches for other AGNs emitting VHE \grs\ have been
conducted using the  air Cerenkov telescopes of the {\it Whipple
Observatory} (Schubnell et al. 1994).  These searches have naturally focused
on those extragalactic $\gamma$-ray sources  already detected at photon
energies $E > 100$ MeV with the EGRET instrument on the {\it  Compton Gamma
Ray Observatory} (Fichtel et al. 1994; Dermer \& Gehrels 1995), but have  so
far yielded negative results. Two explanations have been advanced to explain
these observations. Because Mrk 421 is so close by cosmological standards,
its  detection, coupled with the failure to detect more distant blazars,
implies that TeV \grs\  from the more distant objects are attenuated by the
diffuse intergalactic infrared  radiation field (Stecker, de Jager, \&
Salamon 1992; de Jager, Stecker, \& Salamon 1994).  By contrast, Dermer \&
Schlickeiser (1994) point out that BL Lac is a truly  lineless object (see,
e.g., Kikuchi \& Mikami 1987), and as such has only very tenuous emission
line clouds near the central engine compared to the stong emission lines
found in  quasars (and some BL Lac objects). Thus the rescattered central
source radiation is very  dilute in Mrk 421, permitting the TeV \grs\ to
escape without attenuation.

In 1994 May, Mrk 421 was observed to flare in TeV \grs\ during a period of
$\sim$ 2 days by a factor of 10 over its flux level in the period 1994
Jan.-Apr., or by a factor of 5 above its quiescent flux level measured in
1992 and 1993 (Kerrick et al. 1995; Macomb et al. 1995).  Observations with
{\it
ASCA} made 2 days after the onset of the TeV $\gamma$-ray flare also showed
an enhanced level of 2-10 keV X-rays  compared with earlier {\it EXOSAT}
measurements (Takahashi et al. 1994).  These  observations suggest a method
to diagnose the structure and density of the gas and dust  surrounding the
central source of Mrk 421 and other $\gamma$-ray blazars. By monitoring the
time dependence of the flaring emission from the central source and the pair
attenuation cutoff of the high energy \grs\ due to photon-photon pair
production, we can extract information about the column density, number
density, and location of  the site of high-energy $\gamma$-ray production.
This is a new example of a reverberation  mapping technique which uses
correlated temporal signatures to monitor conditions of a  physical system,
analogous to methods where variability of emission line
intensities are monitored in response to the temporal behavior of photoionizing
radiation from the  central source (cf. Peterson et al. 1991).

In Section 2, we describe the system and model it under the assumption that
the  central source radiation can be approximated by a point source.
Asymptotic forms for the  total photon number density and opacity at times
long after the peak of the flare are derived in Section 3, and numerical
simulations are presented in Section 4. We summarize in Section 5.

\section{Time-dependence of photon opacity from scattered radiation}

Attenuation of MeV-TeV \grs\ by photons with energies in the infrared through
X-ray range can occur in the central regions of active galactic nuclei if
both  the intensity of the lower energy radiation and the scattering depth
of the  surrounding medium are sufficiently great, provided that the spatial
extent of
the scattering medium  is not too large (e.g., Blandford 1993; Dermer
\& Schlickeiser 1994; Blandford \& Levinson 1995). The presence of strong
emission lines from quasars implies the existence of surrounding clouds of
gas which can  rescatter the central-source radiation (Sikora, Begelman, \&
Rees 1994). The structure and  column density of the surrounding gas near
the supermassive black holes powering  quasars is, however, not well known.
Information about the location of the $\gamma$-ray  emission site and the
spatial extent and density of gas and dust surrounding the supermassive
black hole can be extracted using the method we now describe.

We consider the attenuation of \grs\ emitted outward along the z-axis at
height $z_i$ above a central soft-photon source (e. g., an accretion disk)
which
exhibits a flare described, for simplicity, by a Gaussian in time:

$$ \dot N_{ph} (\epsilon, \Omega; t) = \dot N_{max} (\epsilon, \Omega) \>
e^{-{t^2 \over 2 \sigma^2}}\;.
\eqno(1) $$

\noindent In eq. (1), $\dot N_{ph} (\epsilon, \Omega; t) d\epsilon d\Omega$ is
the differential number of photons emitted per second at time $t$ with
dimensionless energy $\epsilon\equiv h\nu/$m$_e$c$^2$ in the range
$\epsilon$ and $\epsilon + d\epsilon$ that are directed into the solid angle
element $d\Omega$  in the direction $\Omega = (\theta,\phi)$.  The full-width
half maximum duration of the flare is given by $2^{3/2}(\ln 2) \sigma \cong
1.96
\sigma$, and the photon emissivity at the peak of the flare is denoted by $\dot
N_{max} (\epsilon, \Omega)$. Some of these soft photons are scattered by
surrounding  clouds of density $n_e (\rv)$ into the path of the \grs.  The
scattered soft photons are directed at an angle $\theta^{\ast} = \cos^{-1}\mus$
with respect to the z-axis, as shown in Figure 1.  We assume that the gas is
moving nonrelativistically and can be treated as a stationary scattering
medium. We also assume that the gas is optically thin to Thomson scattering so
that only single  scatterings are important.

\begin{figure}
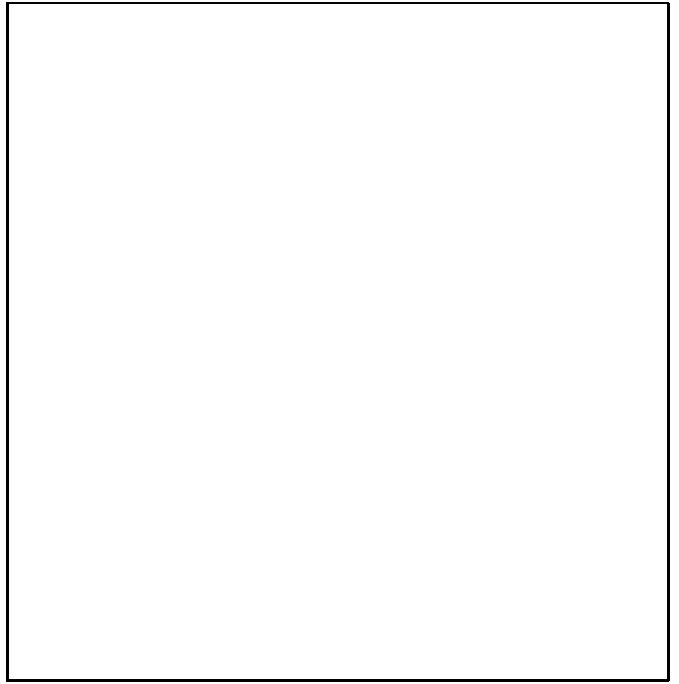

\picplace{9.0 cm}
\caption[ ]{Geometry of the system.  Soft photons are emitted at time
$t$ from an accretion disk which is approximated by a point source at the
origin. The photons are Thomson-scattered at location ${\bf r}$ by electrons
in surrounding dust and gas clouds.   The scattered radiation intercepts
high-energy photons which are beamed outward along the z-axis, and
contribute to the opacity of the high-energy gamma rays through
$\gamma-\gamma$ pair production attenuation. }
\end{figure}

The opacity of a photon with dimensionless energy $\epsilon_1$ due to
photon-photon pair  production can be written as

$$ \tg (\ee, t_i, z_i) =  \int\limits_{z_i}^{\infty} dz \> \int\limits_0^{2\pi}
d\phis\> \int\limits_{-1}^1 d\mus (1-\mus) \int\limits_{2 \over \ee (1 -
\mus)}^{\infty} d\epsilon \cdot $$
$$\cdot  \sg (\ee,\epsilon, \mus) \;n_{ph} (\epsilon, \mus, \phis; z, t_i + {z
-
z_i \over c}). \eqno(2) $$

\noindent (Gould \& Schr\'eder 1967; for corrections see Brown, Mikaelian,
\& Gould 1973). Here $t_i$ denotes the emission time of  the high-energy
photon,
$n_{ph} (\epsilon,\mus,\phis; z, t)$ represents the spectral density of
scattered soft photons per unit  solid angle at height
$z$ and time $t$, and the pair production cross section is given by

$$ \sg (\beta) = {3 \over 16} \st (1 - \beta^2)\cdot $$
$$ \cdot [ (3 - \beta^4) \ln \left({1 + \beta \over 1 - \beta}\right) \, -
\, 2
\beta (2 - \beta^2)]
\eqno(3) $$

\noindent where

$$ \beta = \sqrt{1 - {2 \over \epsilon \ee (1 - \mus)}} \eqno(4) $$

\noindent is the velocity of the electron/positron in the center-of-momentum
frame.

The soft photon density per unit solid angle at location $\rv$ in the direction
$\overline \Omega = (\theta, \phi)$ defined by the location of the scattering
material relative to the position of the soft photon source is given by

$$ n_{ph} (\epsilon, \Omega; {\bf r}, t) = {\dot N_{ph} (\epsilon, \Omega; t -
{r \over c}) \over r^2 \> c} \> \delta (\Omega - \overline\Omega). \eqno(5) $$

\noindent We approximate the Thomson scattering event by isotropic scattering
with cross  section $\sigma_T$ and neglect recoil, which is a good
approximation
when $\epsilon \ll 1$.  The photon  production rate by Thomson scattering at
location {\bf r} and time $t$ is therefore

$$ \dot n_{ph} (\epsilon_s, \Omega_s; {\bf r}, t) = {\st \over 4\pi r^2}\; n_e
({\bf r}) \dot N_{ph} \left(\epsilon_s, \overline\Omega; t - {r \over
c}\right),
\eqno(6) $$

\noindent where the subscript ``s" is used to denote the scattered quantities.

Because the particles at each point in the cloud are assumed to scatter
isotropically, the  spectral density of scattered photons at height $z$ along
the z-axis and at time
$t$ can be  obtained by integrating the scattered spectrum over all space.  We
find that

$$ n_{ph} (\epsilon; z, t) = {\st \over 4\pi c} \int_0^{2\pi} d\phi
\int\limits_{-1}^1 d\mu \int\limits_0^{\infty} dr\; {n_e ({\bf r}) \over x^2}
\cdot $$
$$ \cdot\dot N_{ph} \left(\epsilon, \Omega; t - {r + x \over c} \right) ,
\eqno(7) $$

\noindent where we have dropped the subscript denoting scattered quantities,
and
the overbar on $\Omega$ in the expression for $\dot N_{ph}$.  It is then a
simple matter to obtain the angle-dependence of the photon spectral density
through the expression

$$ n_{ph} (\epsilon, \mus,\phis; z, t) = {\st \over 4\pi c} \int\limits_{-1}^1
d\mu \int\limits_0^{\infty} dr\; {n_e ({\bf r}) \over x^2} \cdot $$
$$ \cdot \dot N_{ph} \left(\epsilon, \Omega; t - {r + x \over c} \right)
\delta (\phis-\bar\phis)\delta\bigl(\mus - \musb [r, \mu, x]\bigr). \eqno(8) $$

\noindent Here $x = (r^2 +z^2-2rz\mu)^{1/2}$ is the distance between the
scattering electron and the z-axis at height z, the angle
$\overline\mus$ is given by

$$ \musb = \sqrt{1 - \left({r \over x}\right)^2 \left(1 -
\mu^2\right)}\;\;,
\eqno(9) $$

\noindent and the angle $\bar\phis = \pi+\phi$ (see Fig. 1). Substituting
equation (8) into equation (2), assuming azimuthal symmetry of the soft-photon
emission spectrum and distribution of scattering material, gives the photon
opacity

$$ \tg (\epsilon_1, t_i, z_i )= {\st \over 2 c} \int\limits_{z_i}^{\infty} dz
\> \int\limits_{-1}^1  d\mus (1-\mus) \int\limits_{2 \over \ee(1 -
\mus)}^{\infty} \! d\epsilon \; \sg (\beta) \cdot $$
$$ \cdot \int\limits_{-1}^1 d\mu \> \int\limits_0^{\infty} dr \;{n_e ({\bf r})
\over x^2} \dot N_{ph} \left(\epsilon, \Omega; t_i + {z - z_i - r - x \over c}
\right) \> \delta\bigl(\mus - \musb \bigr). \eqno(10) $$

In order to perform the $\mus$-integration by evaluating the
$\delta$-function it is necessary to change the order of integrations. The
limits for the remaining integrals then follow from the threshold condition

$$ \epsilon \ee (1 - \musb) > 2 , \eqno(11) $$

\noindent and we find

$$ \tg (\ee, t_i, z_i) = {\st \over 2 c} \int\limits_{z_i}^{\infty} dz \,
\int\limits_{1 \over \ee}^{\infty} d\epsilon \, \int\limits_{-k}^1 d\mu \,
\int\limits_{r_{min}}^{\infty} dr\;{n_e ({\bf r}) \over x^2} (1-\bar\mus)\cdot
$$
$$ \cdot \sg (\epsilon, \ee, \musb) \;\dot N_{ph} \left(\epsilon,
\Omega; t_i + {z  - z_i - r - x \over c} \right)\;, \eqno(12) $$

\noindent where

$$ r_{min} = z \, \sqrt{1 - k^2} \> \bigl( {\mu \, \sqrt{1 - k^2} - k \,
\sqrt{1 - \mu^2} \over \mu^2 - k^2}\bigr) \eqno(13) $$

\noindent if $\mu^2 \ne k^2$, and

$$ k \> = \> 1 - {2 \over \epsilon \, \ee}. \eqno(14) $$

\noindent If $\mu = k$ we have $r_{min} = {z \over 2 k}$ (for this case $k >
0$ is required), while for $\mu = -k$ condition (11) can not be satisfied
($r_{min} \rightarrow \infty$).

To illustrate the importance of the effect of photon attenuation by
reverberating soft photons, we assume that the disk radiates isotropically
and that its radiation during the flare is described by a power law in
photon energy described by the expression

$$ \dot N_{ph} (\epsilon, \Omega; t) = {K \, \epsilon^{-\alpha} \over 4 \pi}
\,  e^{-{t^2 \over 2 \sigma^2}} \;,\eqno(15) $$

\noindent with cutoff energies $\epsilon^l$ and $\epsilon^u$. The power-law
spectra could be produced in the inner hot regions of an accretion disk
by nonthermal electrons which radiate synchrotron photons at frequencies
above the self-absorption frequency.  A more detailed model of the
flaring emission from an accretion disk would require time-dependent solutions
of the accretion disk equations (see, e.g., White \& Lightman 1990), which we
neglect for simplicity.  We also assume for purposes of illustration that the
surrounding clouds are homogeneously distributed with  uniform density
$n_e ({\bf r}) = n_o$.

The constant K is related to the maximum luminosity of the disk by

$$ K = \cases{ {2 - \alpha \over m_e \, c^2} \> {10^{46} L_{46} {\rm
ergs~s}^{-1} \over
\left(\epsilon^u\right)^{2 - \alpha} - \left(\epsilon^l\right)^{2 - \alpha}}
&  for $\alpha \ne 2$
\cr\cr {10^{46} L_{46}{\rm ergs~s}^{-1} \over m_e \, c^2} \ln (\epsilon^u
/\epsilon^l) & for
$\alpha = 2$ \cr}, \eqno(16) $$

\noindent where $L_{46}$ is the maximum disk luminosity in units of
$10^{46}$  ergs s$^{-1}$ and m$_e$c$^2$ is the electron rest mass energy in
ergs.

\section{Asymptotic behavior of photon density and opacity}

If we consider the effects of reverberation at times much later than the
flare  maximum, considerable simplifications of the exact expressions (7)
and (12) for the scattered photon  density and opacity, respectively, are
possible. The derived asymptotic expressions provide useful checks on the
numerical results presented in the next section.

Under the assumptions made at the end of the previous section, we may write the
total scattered photon density as

$$ n_{ph} (\epsilon; z, t) = {n_0 \, \st \, K \, \epsilon^{-\alpha} \over 8
\, \pi \, c \, z} \int\limits_0^{\infty} dr \!\! \int\limits_{\vert r - z
\vert}^{r + z} \!\! dx \> {e^{-{\left(t - {x + r \over c}\right)^2 \over 2
\sigma^2}} \over x \, r }\;. \eqno(17) $$

\noindent At times $ t \gg \sigma $ and $ t \gg {z \over c}$, equation (16)
reduces to

$$ n_{ph} (\epsilon; z, t) \approx {n_0 \, \st \, K \, \epsilon^{-\alpha}
\over  8 \, \pi \, c \, z} \int\limits_0^{\infty} dr \!\!
\int\limits_{-\infty}^{\infty} \!\! dx \> {e^{-{(x - x_0)^2 \over 2 \, \sigma^2
\, c^2}} \over x_0 \, r } \cdot $$
$$ \cdot \Theta(r + z - x_0) \> \Theta(x_0 - \vert r - z \vert)\;, \eqno(18) $$

\noindent where $x_0 = ct - r$ and $\Theta$ denotes the Heaviside function such
that $\Theta(y) = 1$ if $y \geq 0$ and $\Theta(y) = 0 $ otherwise.  Thus, we
have

$$ n_{ph} (\epsilon; z, t) \approx {n_0 \, \st \, K \, \epsilon^{-\alpha}
\over  8 \, \pi \, c \, z} \> \sqrt{2 \, \pi \, \sigma^2 \, c^2} \int\limits_{c
t - z \over 2}^{c t + z  \over 2} { dr \over r} \> {1 \over c t - r} $$
$$ = {n_0 \, \st \, K \, \epsilon^{-\alpha} \over 4 \, \pi \, c \, z} \>
{\sqrt{2 \, \pi \, \sigma^2 \, c^2} \over c t} \> \ln \left({c t + z \over c t
-
z}\right) $$
$$ \approx {n_0 \, \st \, K \, \epsilon^{-\alpha} \over 2 \, \pi } \> {\sqrt{2
\, \pi \, \sigma^2 } \over (c t)^2} \;.\eqno(19) $$

\noindent Note that this last expression  does not explicitly depend on $z$.

Inserting this expression into eq. (2), we set $t = t_i + {z - z_i
\over  c}$, noting that approximation (19) is valid for $c t \gg z_i$. The
disk photon density at the location of the scattering electrons decreases as
$1\over r^2$ and the scattered photon density, again, decreases as
$1 \over x_0^2$. Therefore, the main contribution of scattered photons will
come from that angle $\theta^{\ast}$ with respect to the $z$-axis for which
$x_0 = r$. This yields

$$ \mus_0 = {z \over c t_i + z - z_i}. \eqno(20) $$

Using the scattered photon density

$$ n_{ph} (\epsilon, \mus,\phis; z, t) \approx (2\pi)^{-1}n_{ph} (\epsilon;
z, t)\delta(\mus - \mus_0)\;, \eqno(21) $$

\noindent we find

$$ \tg (\ee, t_i) \cong {n_0 K \st \sigma \over \, \sqrt{2 \pi}} \> \left(c t_i
- z_i\right) \cdot $$
$$ \cdot \int\limits_{z_i}^{\infty} {dz \over (c t_i + z - z_i)^3}
\int\limits_{2 \over \ee (1 - \mus_0)}^{\epsilon^u} d\epsilon \>
\epsilon^{-\alpha} \> \sg (\ee, \epsilon, \mus_0) $$
$$ = {2n_0 K \st \sigma \over \sqrt{2 \pi}} \, \left({\ee \over
2}\right)^{\alpha - 1} \int\limits_{z_i}^{\infty} {dz \over (c t_i + z -
z_i)^2}
\cdot $$
$$ \cdot (1 - \mus_0)^{\alpha} \> \int\limits_{\beta_{min}}^{\beta_{max}}
d\beta \> \beta \, (1 - \beta^2)^{\alpha - 2} \, \sg (\beta) \eqno(22) $$

\noindent where

$$ \beta_{min} = \cases{\sqrt{1 - {2 \over \ee \epsilon^l (1 - \mus_0)}} & if
$\epsilon^l > {2 \over
\ee (1 - \mus_0)}$ \cr 0 & otherwise }\;, \eqno(23 a) $$
\noindent and
$$ \beta_{max} = \sqrt{1 - {2 \over \ee \epsilon^u(1 - \mus_0)}}\;. \eqno(23 b)
$$

\noindent Here we may assume $\beta_{min} \approx 0$, $\beta_{max} \approx 1$
for the model we consider, because the pair production cross section (3)
vanishes for $\beta \to 0$ and for
$\beta \to 1$.

The $z$-integration in eq. (22) can easily be done analytically, and the second
integration is only dependent on the spectral index $\alpha$. Thus, we define

$$ I(\alpha) \equiv \int\limits_0^1 d\beta \> \beta \, (1 - \beta^2)^{\alpha -
2} \, \sg (\beta)\eqno(24) $$
$$ I(1.5) \approx 0.3 \> \st $$

\noindent The function $I(\alpha)$ is only weakly dependent on $\alpha$. The
photon opacity for times late after the flare is given by

$$ \tg (\ee, t_i) = {2^{1/2} \, n_0 \, K \, \st \, \sigma \over \pi^{1/2}} \>
{1
\over \alpha + 1} \> \left({\ee \over 2}\right)^{\alpha - 1} {1 \over c t_i} \;
I(\alpha) \;,\eqno(25) $$

\noindent which is independent of $z_i$ because we neglected $z_i$ compared to
$c t_i$ in equation (25).

\section{Numerical Results}

We now show the results of numerically performing the integrals (7) and (12),
and compare these results with approximations (19) and (25), respectively,
derived in Section 3.  We assume that the medium surrounding the central engine
has a Thomson scattering optical depth of $0.02$, extends out to a scale height
of $0.1$ pc, and is uniformly distributed in this region. This yields an
electron density $n_0 = 10^5$ cm$^{-3}$. Furthermore, we describe the disk
photon spectrum during the flare by a power law with spectral index
$\alpha = 1.5$ extending from dimensionless photon energy $\epsilon^l =
10^{-8}$
(corresponding to $5 \cdot 10^{-3} \>{\rm eV}$) to $\epsilon^u = 10^{-2}$ \quad
($5
\cdot 10^3 \> {\rm eV}$), and assume its maximum luminosity $L_{max} = 10^{46}$
ergs s$^{-1}$.

\begin{figure}
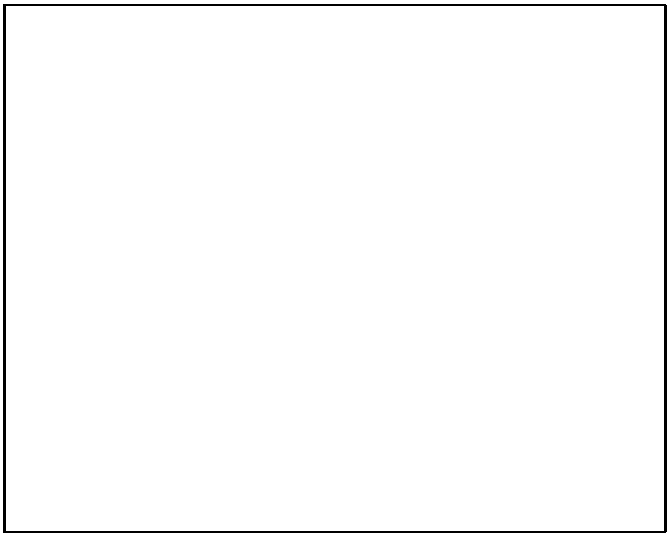

\picplace{7.0 cm}
\caption[ ]{Time dependence of the scattered photon number density and
pair-production opacity $\tg$.  The photon number density is computed at
$z=10^{-2}$ pc, and $\tg$ is for a 500 GeV photon ($\epsilon_1 = 10^6$)
interacting with lower-energy photons produced in an accretion-disk flare
described by a Gaussian in time with a full-width half-maximum duration of one
week. The approximations valid at late time are shown by the dotted curves. }
\end{figure}

The time-dependent effect of a one-week flare described by equation (15) on
the pair-production opacity for 500 GeV photons
($\epsilon_1 = 10^6$) emitted at $z_i = 10^{-2}$ pc, together with the
total scattered-photon number density at the injection height, is shown in
Fig. 2. We see that the photon opacity reaches maximum values of order unity
and, due to the finite light travel time, attains its maximum at times later
than at $t = z_i / c$ which, for the example considered here, is $\sim 10^6
\>$ s. The numerical results show good agreement with the approximations
for the asymptotic behavior at late times, which indicates a decrease of the
opacity as $1/t$. If $c t > 0.1$ pc ($t
\gte 10^7$ s), of course, the numerical results will decrease faster than
the approximation because the latter does not take into account the lack of
scattering material at $r > 0.1$ pc.

\begin{figure}
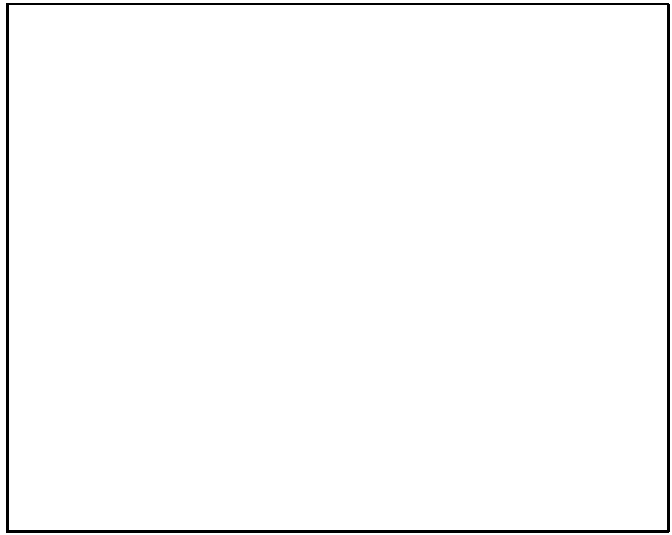

\picplace{7.0 cm}
\caption[ ]{Time dependence of the
pair-production opacity $\tg$ for photons emitted at different heights
$z_i$ given by the labels on the curves.  Other parameters are the same as
in Fig. 2. The dotted curve shows approximation (25) for $\tg$.}
\end{figure}

In Fig. 3 we show the dependence of the photon opacity on the injection
height $z_i$, with all other parameters the same as given in the case
shown in Fig. 2. From variability time scale arguments, Kerrick et al. (1995)
derived an upper limit of 0.03 pc on the size of the emitting region of the TeV
radiation for the case of Mrk~421. This is in accord with the values
of $z_i$ used in Fig. 3 if the injection height and inferred size scale are
comparable. As the injection height increases, the maximum photon opacity
decreases and attains its greatest value at increasingly later times. It is
interesting to note that, as we already saw from equation (25), the photon
opacity late after the flare maximum does not depend on the injection height.
Note also that the delay between $t = z_i / c$ and the time of maximum opacity
depends very weakly on the injection height. In the case considered
here, the delay $\Delta t \approx 10^6$ s, corresponding to the duration of
the flare.

\begin{figure}
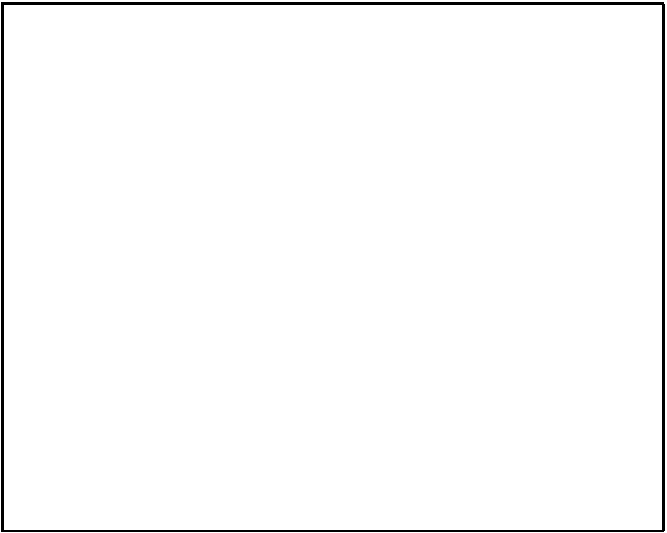

\picplace{7.0 cm}
\caption[ ]{Time dependence of the
pair-production opacity $\tg$ for photons emitted at height
$z_i = 10^{-3}$ pc with dimensionless
energies $\epsilon_i$ given by the labels on the curves.  Other
parameters are the same as in Fig. 2. The dotted curves show approximation
(25) for $\tg$.}
\end{figure}

The energy dependence of the opacity $\tg$ is shown in Fig. 4 where, again,
the dotted curves represent the analytic approximation (25). For these
calculations, we assumed an injection height of $z_i = 10^{-3}$ pc. From this
figure, we see that the form of the time-dependent photon opacity is
esentially the same at all gamma-ray energies.  The relative amplitude of
the opacity curves is described by

$$ \tg (\ee, t_i) \propto \ee^{\alpha - 1}\;, \eqno(26) $$

\noindent as was shown for the asymptotic behavior (eq. [25]).

\begin{figure}
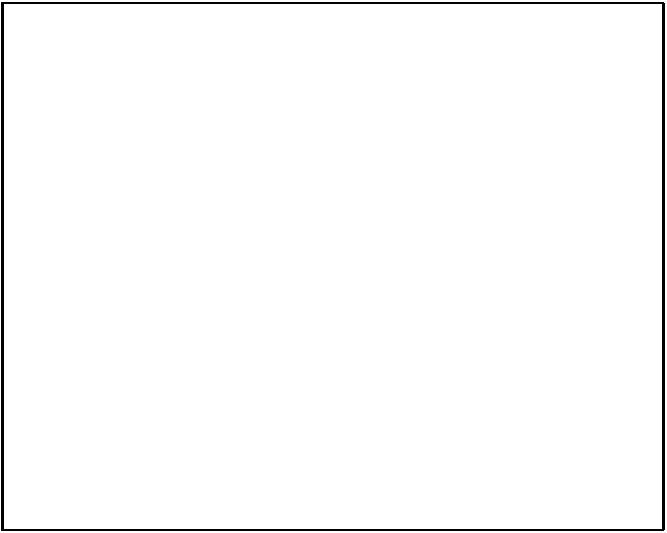

\picplace{7.0 cm}
\caption[ ]{Time dependence of the observed photon spectrum due to
photon-photon attenuation by soft photons from an accretion-disk flare
which are scattered into the high-energy photon beam. Here we assume a
power-law photon spectrum with photon spectral index $\alpha = 2$.  The
high-energy photons are assumed to be emitted at height $z_i = 10^{-3}$ pc.
Other parameters are the same as in Fig. 2. The curves are labeled by time
after flare maximum.  }
\end{figure}

Fig. 5 shows the time-dependent modification of a primary
$\gamma$-ray spectrum by the effects of pair-production opacity effects from
accretion disk-flare photons. Here we let the primary gamma-ray
spectrum be described by a power law with spectral index $\alpha = 2.0$, and
choose an emission height $z_i = 10^{-3}$ pc. The results are plotted for
different times with respect to the time of the maximum of the
accretion-disk flare.  As can be seen,
$\gamma$-rays with energies
$\gte 500$ GeV will be almost completely absorbed after $\sim 10^4$ s following
the flare maximum and will remain strongly attenuated long after the flare. If
we
assume a greater injection height, this effect will occur correspondingly
later and the influence of reverberating photons will be restricted to
higher energy
$\gamma$-rays, as can be seen from Fig. 4.

\section{Discussion and Summary}

We have considered a highly simplified geometry to illustrate the
potential of gamma-ray astronomy to map the structure and density of
clouds of dust and gas surrounding the central nucleus of a BL Lac object
or quasar. Photon-photon
attenuation intrinsic to the source will occur only if the surrounding
medium has a sufficiently large column density or if the gamma-ray emission
site is sufficiently near the central nucleus. As pointed out in the
Introduction, the scattering emission line clouds
are relatively tenuous in Mrk 421 compared to quasars. It is therefore
likely that the number density of the scattering medium in Mrk 421 is
significantly less than
$10^5$ cm$^{-3}$, the value assumed in our  model calculations. As the pair
production optical depth scales roughly  linearly with the number density of
the
scattering material, this is  consistent with the observational result by
Kerrick et al. (1995), who did  not find any temporal variation of the spectral
shape during the TeV $\gamma$-ray flare in 1994 May.

This observation implies a limit on the density of the
scattering material surrounding the central engine of Mrk~421 if the TeV
$\gamma$-ray production site is within 0.03 pc (see Section 4).
Using  Eq.~(26) and the results shown in Fig. 3, and assuming that the pair
production optical depth up to several TeV ($\ee \sim 10^7$) always remained
less
than 0.1, we find that

$$ n_0 \lte  10^4 \> \ee^{1 - \alpha} \, L_{46}^{-1} \> {\rm cm}^{-3} \approx 3
\> L_{46}^{-1} \> {\rm cm}^{-3}
\eqno(27)
$$

\noindent for a Gaussian flare with a FWHM duration of 1 week. Optical
and TeV gamma-ray campaigns to monitor Mrk 421 can therefore provide
constraints on the density of the surrounding medium of Mrk 421. During the May
1994 outburst of Mrk 421, however, only variations in the keV and TeV ranges
were observed, whereas no significant variations were seen in
near-simultaneous mm, IR, UV, and GeV gamma-ray observations (Macomb et al.
1995). The keV X-rays are most effective in attenuating 0.1- 1 GeV gamma rays.
We could apply our results to this regime, noting that the X-ray flare
luminosity was at the level of $\sim 2\cdot 10^{44}$ ergs s$^{-1}$. Although
the poor statistics of GeV observations might not support a constraint as
strong as $\tg \lte 0.1$, equation (27) should yield a sensible estimate for
this case. This implies a density of $n_0 \lte 10^4$ cm$^{-3}$ for
a one week X-ray flare, corresponding to a Thomson scattering optical depth of
$\tau_T \lte 0.002$ of the surrounding scattering medium if it exends out to
a scale height of 0.1~pc. Accounting for probable inhomogeneities (e. g., a
decrease of the cloud density with distance from the center) together with the
relatively high location of the $\gamma$-ray emission site, this limit on the
Thomson scattering optical depth should be raised somewhat in order to yield a
more conservative estimate. It should be kept in mind, however, that the X-ray
flare in this event is probably beamed emission from the jet, in which case
these estimates do not apply.

As shown in Fig. 5, the time-dependent truncation of the gamma-ray spectra
following an accretion-disk flare gives useful information about the location
of the gamma-ray emission region and the mean density of the surrounding gas
and dust.  By comparing the temporal behavior of the flare and the attenuated
gamma-ray spectrum in detail, further information about the distribution of
scattering gas can be obtained.

We can predict idealized
time-dependent behavior of $\gamma$-ray emission in response to variations
in the lower-energy radiation for specific models.  In the models of Dermer
\& Schlickeiser (1993) and Sikora et al. (1994), gamma-ray production
in blazars is caused by Compton scattering of accretion-disk photons or
rescattered radiation. Flaring accretion-disk radiation in the infrared to
X-ray
range will produce flaring behavior at 100 MeV to GeV energies due to the
time-variable soft photon energy density in the relativistically outflowing
fluid frame. If there are extremely energetic electrons in the fluid
frame, then a flare can also occur in the TeV range. The shape of the
unabsorbed IC-spectrum has been calculated for isotropic relativistic
electrons in the outflowing fluid frame  (Dermer \& Schlickeiser 1993). If
the accretion-disk flaring behavior in the infrared and optical range
shows the same behavior as the UV/X-ray flare, then the
$\gamma$-ray spectrum will be absorbed according to the time and energy
dependence illustrated in Fig. 5.

It is important to note that it is the infrared and optical photons that are
most effective in attenuating TeV gamma rays (e.g., Stecker et al. 1992).
To apply our method to TeV measurements, therefore, correlated
infrared/optical and high-energy gamma-ray observations are necessary (note
that no space-based detectors may be necessary in this case).   However, the
diffuse intergalactic infrared background radiation produced by the
superposition of reprocessed emission from stellar nucleosynthesis renders most
quasars with redshifts $\gte 0.2$ invisible, due to strong
$\gamma-\gamma$ attenuation over this pathlength (see also Stecker \& de
Jager 1993; MacMinn \& Primack 1994).  For applications to a larger
number of objects, it is therefore desirable to devise $\gamma$-ray detection
systems in the 10 GeV - several hundred GeV regime, where the
effects of the diffuse infrared background radiation are small.  The
identification of attenuation intrinsic to the source will be necessary, in
any case, to implement methods to infer the Hubble constant from
high-energy gamma-ray observations (Salamon, Stecker, \& De Jager (1994)).

Finally, we note that the presence of scattered radiation in the source
will confuse any attempt to determine the level of the intergalactic infrared
background radiation using high energy $\gamma$-ray observations
of blazars (Stecker et al. 1992; Dermer \& Schlickeiser 1994).  By
monitoring temporal variations at low and high-photon energies, our method
provides a way to subtract out the level of attenuation
intrinsic to the source, and determine whether the diffuse intergalactic
infrared background or rescattered radiation is more important for
attenuating high-energy photons in a given source.

\acknowledgements{We thank Dr. R. Schlickeiser for comments and discussions.
Useful comments by Dr. John Mattox and the referee, Professor O. C. de Jager,
are also acknowledged. Partial support for the work of C. D. was provided
under NASA grant DPR S-30931-F. M. B\"ottcher acknoledges financial support
by the Deutsche Forschungsgemeinschaft.}

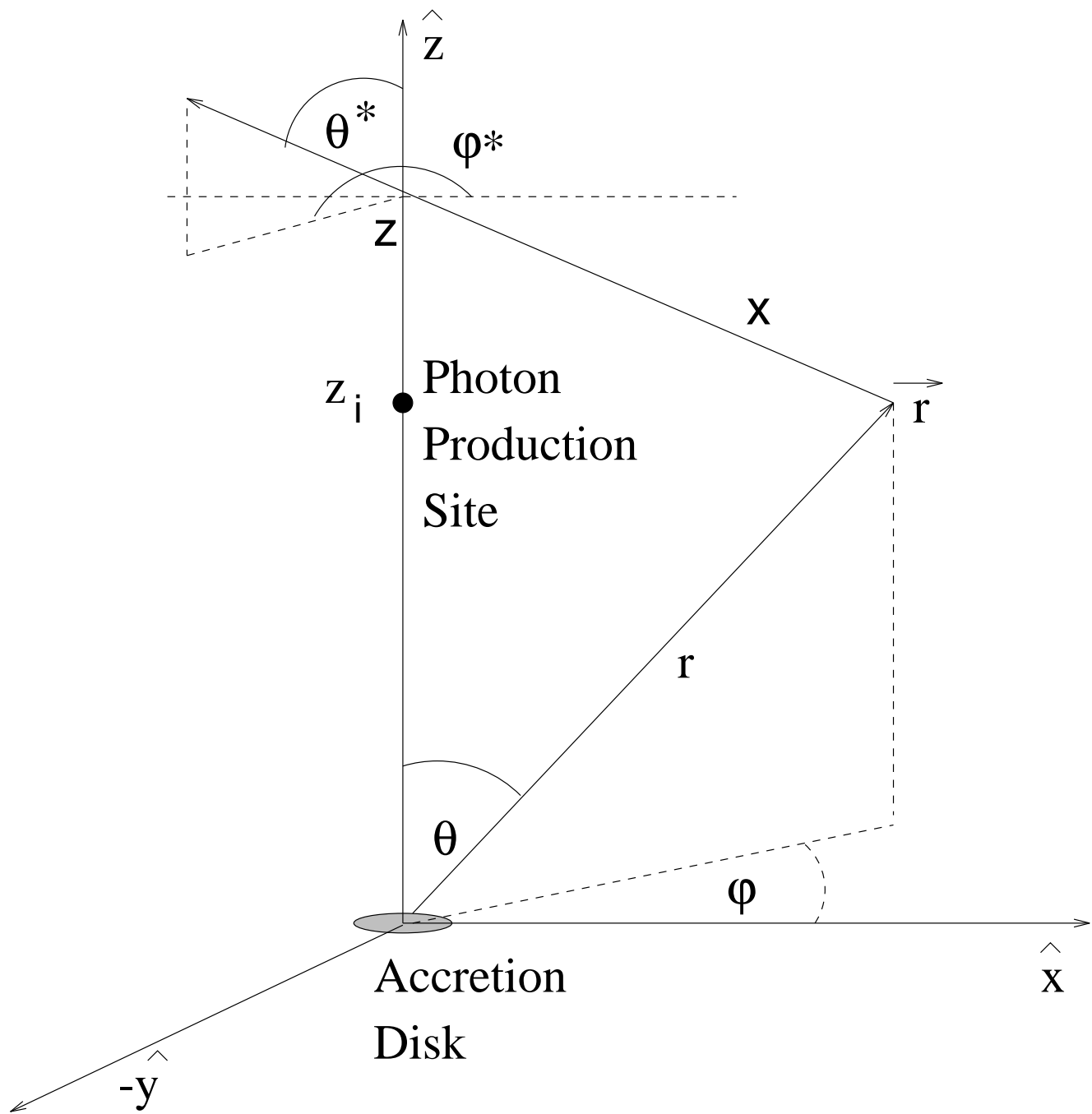
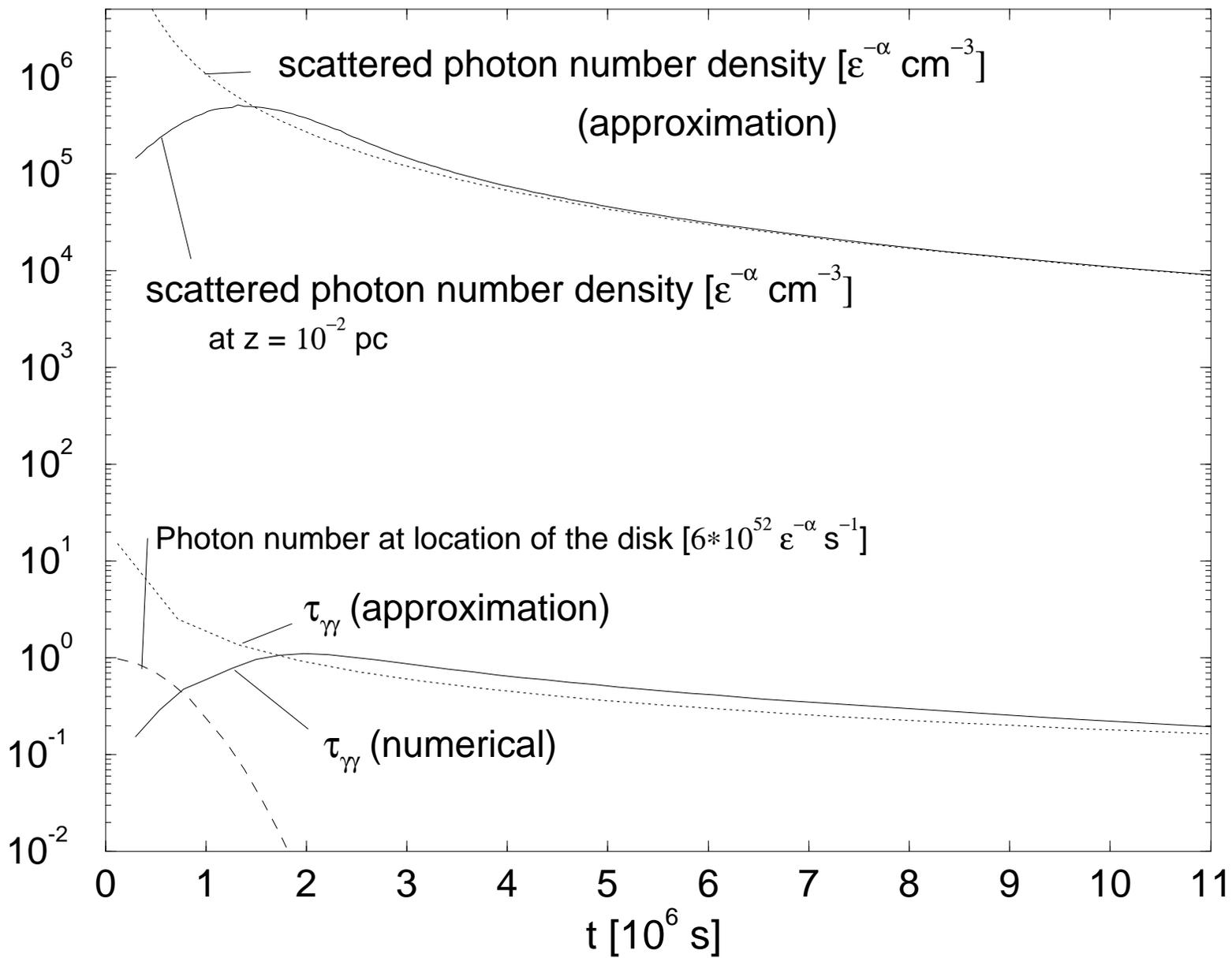
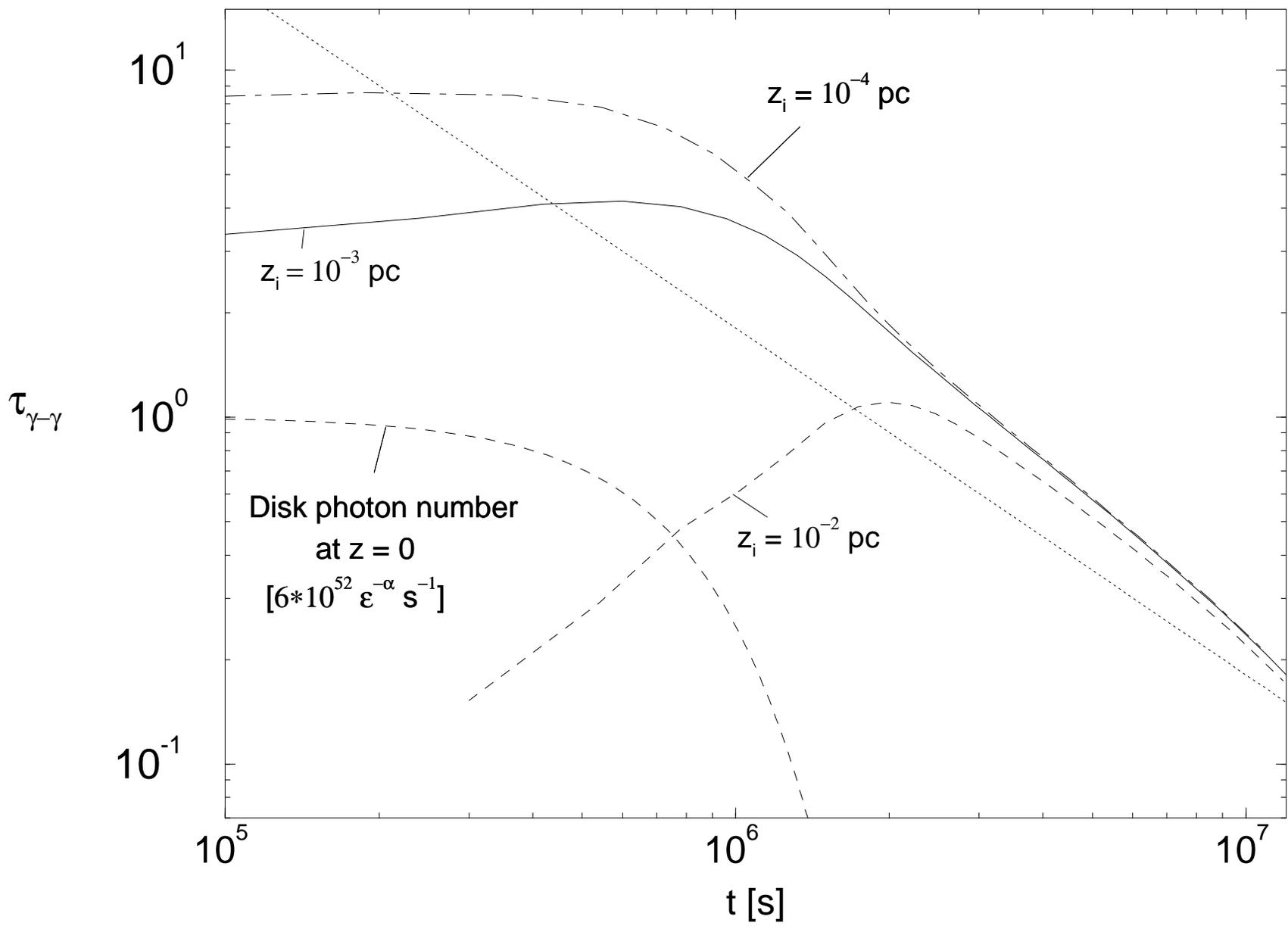
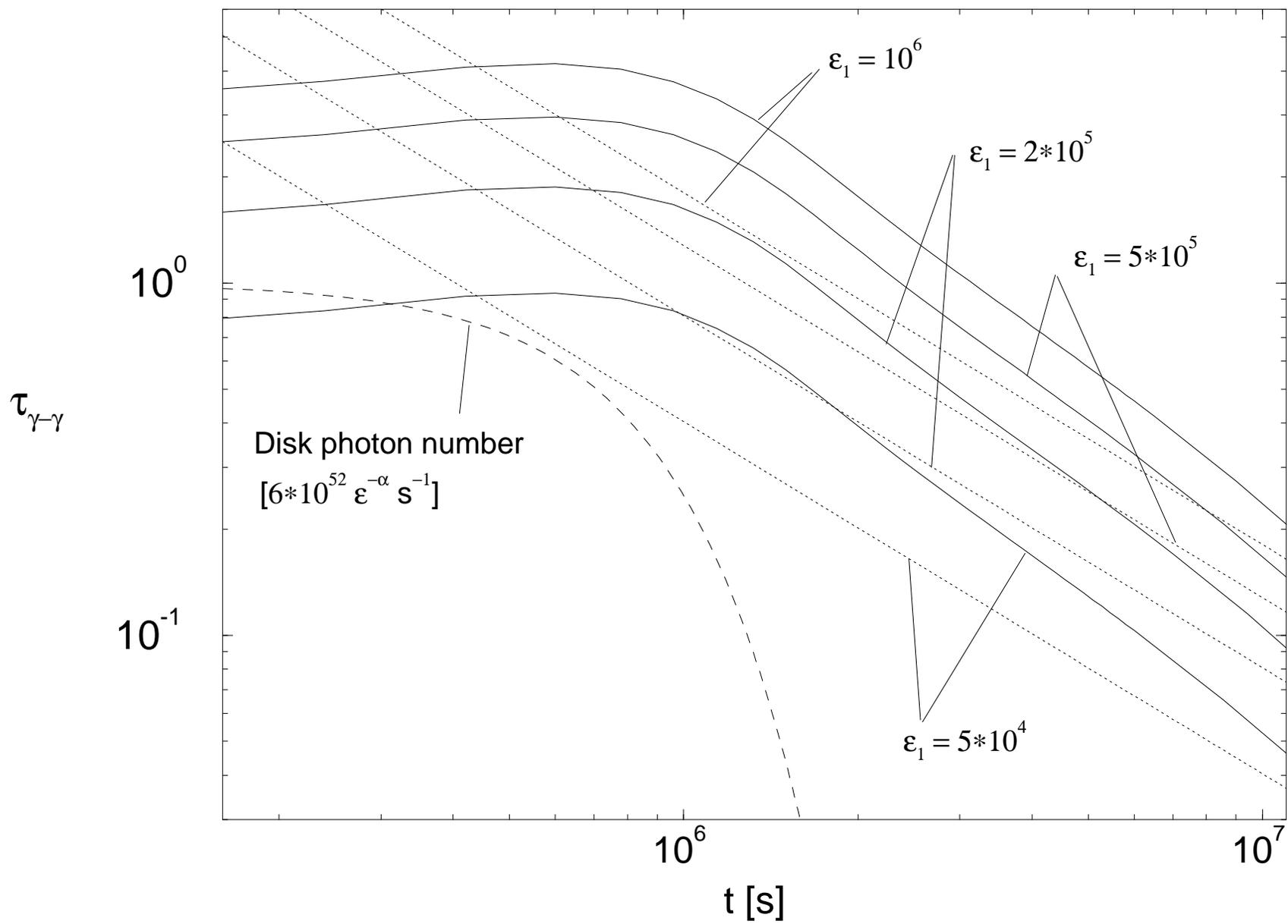
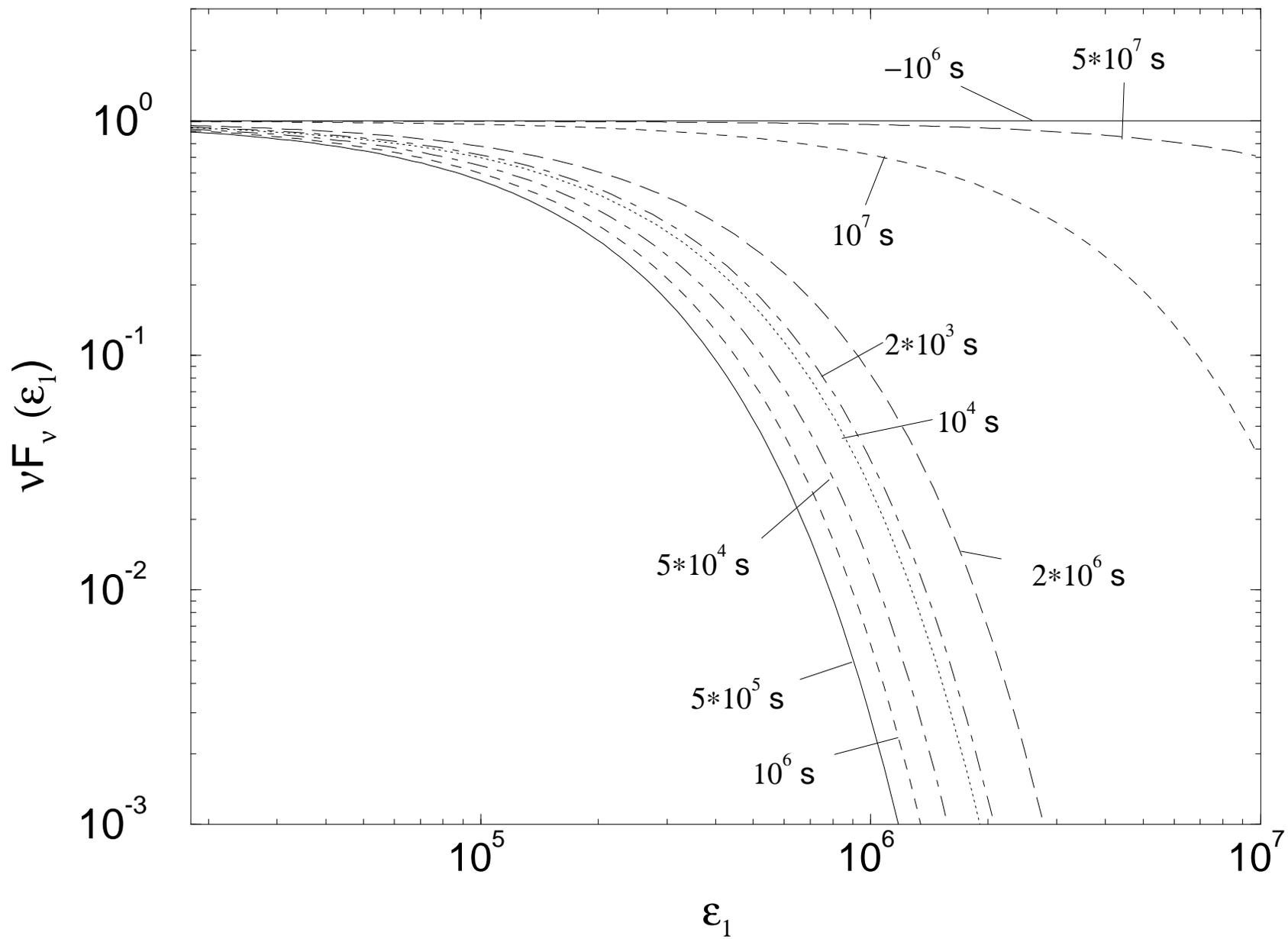
\end{document}